# Efficient current-induced spin torques and field-free magnetization switching in a room-temperature van der Waals magnet

Chao Yun[1,2,3*†], Haoran Guo[1†], Zhongchong Lin[1], Licong Peng[2], Zhongyu Liang[1], Miao Meng[4], Biao Zhang[2], Zijing Zhao[2], Leran Wang[1], Yifei Ma[4], Yajing Liu[3], Weiwei Li[3], Shuai Ning[4], Yanglong Hou[2*], Jinbo Yang[1*], Zhaochu Luo[1*]

[1]State Key Laboratory for Mesoscopic Physics, School of Physics, Peking University, Beijing 100871, China
[2]School of Materials Science and Engineering, Peking University, Beijing 100871, China
[3]MIIT Key Laboratory of Aerospace Information Materials and Physics, State Key Laboratory of Mechanics and Control of Mechanical Structures, College of Physics, Nanjing University of Aeronautics and Astronautics, Nanjing 211106, China
[4]Tianjin Key Lab for Rare Earth Materials and Applications, Center for Rare Earth and Inorganic Functional Materials, School of Materials Science and Engineering, Nankai University, Tianjin 300350, China

*Corresponding author email: yunchao@pku.edu.cn (C.Y.); hou@pku.edu.cn (Y.H.); jbyang@pku.edu.cn (J.Y.); zhaochu.luo@pku.edu.cn (Z.Luo)

† C.Y. and H. G. contributed equally to this work.

**Abstract:** The discovery of magnetism in van der Waals (vdW) materials has established unique building blocks for the research of emergent spintronic phenomena. In particular, owing to their intrinsically clean surface without dangling bonds, the vdW magnets hold the potential to construct a superior interface that allows for efficient electrical manipulation of magnetism. Despite several attempts in this direction, it usually requires a cryogenic condition and the assistance of external magnetic fields, which is detrimental to the real application. Here, we fabricate heterostructures based on $Fe_3GaTe_2$ flakes that possess room-temperature ferromagnetism with excellent perpendicular magnetic anisotropy. The current-driven non-reciprocal modulation of coercive fields reveals a high spin-torque efficiency in the $Fe_3GaTe_2$/Pt heterostructures, which further leads to a full magnetization switching by current. Moreover, we demonstrate the field-free magnetization switching resulting from out-of-plane polarized spin currents by asymmetric geometry design. Our work could expedite the development of efficient vdW spintronic logic, memory and neuromorphic computing devices.

## Introduction

Two-dimensional (2D) van der Waals (vdW) magnets have recently emerged as one of the most intriguing areas in physics and materials research. These cleavable magnetic materials provide an unprecedented platform to manipulate nanoscale phases and to host emergent spin-related phenomena, thus promising appealing applications in low-power electronics, quantum computing and optical communications (*1–6*). Owing to their intrinsically clean surface without dangling bonds, the vdW magnets have the potential to realize a superior interface that allows for efficient spin transports. Several high-quality crystals of vdW magnets have been synthesized and mechanically cleavable down to a thickness of a few nanometers, where the long-range magnetic order and particularly, the strong perpendicular magnetic anisotropy (PMA), can still be sustained (*7–12*). These excellent capabilities enable versatile manipulation of magnetic order by current-induced spin torques (STs) on the basis of efficient charge-to-spin conversion. Especially, current-induced magnetization switching of vdW magnets has been achieved in the heterostructures incorporating the materials with strong spin-orbit coupling effect such as heavy metals (Pt, Ta and W) (*13–17*), transition-metal dichalcogenides (WTe$_2$) (*18-20*) and topological insulators (($Bi_{1-x}Sb_x)_2Te_3$) (*21,22*), which offers the elemental functionality to construct vdW magnet-based memory devices. Despite these successful attempts in electrical manipulation of magnetization in vdW magnets, it usually requires a cryogenic condition and the assistance of external magnetic fields, which hinders its real application. Moreover, the full magnetization switching of 100% volume fraction in vdW magnets which is essential to guarantee a reliable operation in a downscaled element, remains elusive. This undesired performance has been ascribed to several reasons such as the non-uniformity of magnetic domain nucleation (*23*), current-induced demagnetization due to Joule heating (*13*), geometry-induced domain-wall pinning (*24*) and interface contamination during the device fabrication (*15*). Recently, a room-temperature vdW ferromagnet Fe$_3$GaTe$_2$ (FGT) has been reported to possess a high Curie temperature (350-380 K) and large PMA (4.8×10$^5$ J/cm$^3$) (*25*). In addition, large tunneling magnetoresistance (up to 85% at room temperature) (*26,27*) and current-induced magnetization switching have been observed in the FGT-based heterostructures (*28*), bringing vdW spintronic devices one step closer to the application.

In this work, we fabricated a high-quality FGT/heavy metal heterostructure and investigated the effect of current-induced STs. The current-induced STs can facilitate/prohibit the magnetic domain-wall motion in FGT, resulting in a non-reciprocal modulation of coercive fields with a high ST efficiency (45.5 Oe MA$^{-1}$cm$^2$). We then achieved a full magnetization switching at a relatively low switching current density (4.8 MA cm$^{-2}$) at room temperature. Moreover, we designed the asymmetric geometry by covering one side edge of the FGT flake with Pt and observed the field-free magnetization switching operation due to the injection of additional out-of-plane polarized spin currents. Our work reveals the possibility of FGT/heavy metal heterostructure in the applications of low-consumption, reliable and scalable spintronic devices.

**Results and discussions**

Single crystals of FGT with the size of ~2 × 2 × 0.2 mm (length × width × thickness) were prepared by self-flux method (details in the Methods section). The vdW atomic structure of FGT, as shown in Fig. 1A, is constructed with Fe$_3$Ga herometallic slabs sandwiched between two Te layers and there is a vdW gap between the adjacent Fe$_3$GaTe$_2$ layers. The X-ray diffraction (XRD) pattern of the as-grown bulk crystal only contains FGT (0 0 $l$) peaks with a narrow full width at half maximum (FWHM < 0.02°) (Fig. 1B), indicating the high-quality single crystallinity. Moreover, Raman spectra only exhibit characteristic peaks of lattice vibration modes A$_1$ (~132 cm$^{-1}$) and E$_2$ (~154 cm$^{-1}$) of Te atoms (Fig. S1), which are the feature of 2D telluride compounds (*29*). We then conducted the magnetization measurements using vibrating-sample magnetometer (VSM) to study the magnetic properties of the FGT bulk. The magnetization hysteresis loops with the magnetic fields parallel and perpendicular to the *c*-axis of FGT crystal reveal that the FGT bulk has an easy axis along *c*-axis with a high effective anisotropy field of ~32 kOe at room temperature (Fig. 1C). This ferromagnetic order with perpendicular magnetic anisotropy sustains up to 360 K (Fig. 1D), which guarantees the potential room-temperature application.

In order to fabricate vdW spintronic devices, FGT flakes with the thickness below 20 nm were mechanically exfoliated with scotch tape or blue membrane tapes and then transferred onto a 200 nm-thick SiO$_2$ layer on a silicon substrate in the glovebox enclosure with inert gas atmosphere. As shown in Fig. 1E, the optical microscopic image reveals the microscopic morphology of FGT flakes and atomic force microscopy (AFM) depth profile indicates their thickness of ~15 nm (Fig. S2). Then we sputtered a 7 nm-thick Pt layer on exfoliated FGT flakes with a low deposition power (20 W) to mitigate the bombardment damage of Pt atoms on the FGT surface. The FGT/Pt heterostructure was further patterned into Hall bar structures for electrical measurements using photolithography combining the ion milling process. As shown in the cross-sectional high-resolution transmission electron microscopy (HR-TEM) image (Fig. 1F), the layered lattices in FGT flakes indicate high-quality interface between FGT and Pt and confirm the thin thickness of exfoliated FGT flakes. No obvious oxidation or degradation is observed in FGT and at the FGT/Pt interfaces. The high-quality vdW magnet flake and the clean/sharp interface in the heterostructure are crucial to achieving the efficient manipulation of magnetization by spin currents.

To study the magnetic properties of the FGT/Pt heterostructure, we first quantify the magnetic anisotropy and Curie temperature of the FGT flakes by exploiting the anomalous hall effect (AHE) with a relatively low current of 0.5 mA passing through the heterostructure (Fig. 1G). When the magnetic field is applied along the out-of-plane direction (*c*-axis), the transverse resistance ($R_{xy}$) hysteresis loop exhibits a square shape with a coercive field ($H_C$) of 590 Oe (inset of Fig. 1G), indicating an excellent perpendicular magnetic anisotropy at room temperature. When the magnetic field is in-plane, $R_{xy}$ gradually changes to the middle value with the increase of the magnetic field, as the magnetization is aligned into the plane and leads to the vanishing of $R_{xy}$. By fitting the in-plane hysteresis loop, we can obtain the effective magnetic anisotropy field $H_K$ = 30.0 kOe, corresponding to the magnetic anisotropy energy $E_{an} = H_K M_S/2 = 6.36\times10^5$ J/m$^3$. Note that the strong perpendicular magnetic anisotropy allows the downscaled size of FGT-based storage unit to reach 2.0 nm while the thermal stability factor $E_{an}/k_B T$ remaining more than 40, which is highly desired for high-density magnetic memory. With the increase of temperature, the coercive field decrease as well as the magnitude of AHE (Fig. 1H). $R_{xy}$ vanishes above 350 K, giving the high Curie temperature of the FGT flake (Fig. 1I). Both magnetic anisotropy and Curie temperature of FGT flakes are very close to that of the FGT bulk, indicating that the magnetic properties have been well preserved after the fabrication process of being exfoliated into nanometer-thick flakes, constructed to heterostructures and patterned to Hall devices.

To reveal the influence of current-induced STs, we measured the AHE hysteresis loops by applying electric currents with different magnitudes. As shown in Fig. 2A, with the increase of applied currents, the coercive field decreases gradually as the electric current-induced Joule heating leads to an elevated temperature in FGT. Compared to temperature-dependent AHE hysteresis loops measured with low currents (Fig. 1H), the electric current-induced temperature change is less than 8 K (at 7.5 mA). Interestingly, the decrease of the coercive field is non-reciprocal with respect to the polarity of electric currents (Fig. 2B): the coercive field decreases almost linearly with the increase of positive currents, while for negative currents, there is a plateau of the coercive field occurring when $I$ > -5 mA and a rapid decrease when $I$ < -5.5 mA. This current-induced non-reciprocal modulation of coercive fields distinctly deviates from the effect caused by the pure Joule heating that should be symmetric for positive and negative currents. Indeed, this phenomenon is similar to the magnetism modulation recently reported in a magnetic Weyl semimetal Co$_3$Sn$_2$S$_2$ that is attributed to the ST-driven domain-wall motion (30). As illustrated in Fig. 2C, we consider an elongated exfoliated FGT flake with a domain nucleation centre located at one end of the structure. During the hysteresis measurement with the initial magnetization ⊙, a small ⊗ domain nucleates first at the bottom nucleation centre at small magnetic field pointing to ⊗. The ⊗|⊙ domain wall then propagates from bottom to top with the increase of magnetic fields, accomplishing the switching of the whole flake. The current-driven STs can facilitate or hinder the domain-wall motion depending on the polarity of the electric current. When the STs facilitate the domain-wall motion, both the magnetic field and STs drive the domain-wall propagate in the same direction, leading to magnetization switching at low magnetic fields and hence a small coercive field. By contrast, when the STs hinder the domain-wall motion, it requires a high magnetic field to overcome the impediment of the STs, manifesting with large coercive fields in the hysteresis loop.

We could estimate the efficiency of the ST-induced magnetism modulation effect by calculating the ratio between the coercive field change and applied current density $\Delta H/\Delta J$. As shown in Fig. 2D, in order to exclude the coercive field change caused by the Joule heating effect, we extract the change that is induced by the pure thermal effect by using the longitude resistance $R_{xx}$ vs. temperature relation as a internal thermometer (details in Supplementary Information S3). The magnitude of the ST efficiency reaches 45.5 Oe MA$^{-1}$cm$^2$, which is higher than typical metallic ferromagnetic systems (*31,32*), indicating the potential of FGT to serve as a novel vdW candidate for highly-efficient spintronics. Note that the source of the efficient current-induced ST may come from the spin-transfer torque (STT) effect associated within the domain wall in magnets (*30,31*) or spin-orbit torque (SOT) effect originating from the vdW magnet bulk (*33,34*) and FGT/Pt interface (*13,14*).

To further support the microscopic mechanism of ST-driven domain-wall motion, we recorded the hysteresis loops with the magnetic field rotating in the *x-z* plane by applying a relatively small current of 0.5 mA (Fig. 2E). The change of coercive fields as a function of tilt angle $\theta$ is fitted well with the relation of 1/cos $\theta$, implying that the dominant magnetic reversal mechanism is domain-wall motion rather than coherent rotation (*35*). This confirms that the observed non-reciprocal modulation of coercive fields can be correlated to domain nucleation and current-assisted domain-wall propagation.

We have demonstrated efficient modulation of coercive fields via current-induced STs. Another appealing performance of the manipulation of magnetism is to switch the magnetization with electric currents, which is the cornerstone of magnetic random-access memory (MRAM) devices. Despite several reports on the current-induced magnetization switching in vdW magnets, the performance usually deviates from the full switching scheme, arising reliability concerns about the applications of downscaled vdW memory devices. The partial switching performance has been attributed to various underlying mechanisms such as Joule heating (*13*), geometry-induced domain-wall pinning (*24*) and interface contamination during the device fabrication (*15*).

In experiment, we grew the Pt layer at low deposition power to eliminate the bombard damage at the FGT/Pt interface and fabricated the pillar-like structure which can further get rid of the shunt effect of the Hall bar geometry (Fig. 3A). The sharp switching behavior with out-of-plane magnetic field indicates excellent perpendicular magnetic anisotropy and low pinning effect (Fig. 3B). We then measure the current-induced magnetization switching performance with an in-plane magnetic field that breaks the symmetry. During the measurement, a short current pulse with the width of 1 ms is applied to switch the magnetization, and a followed-up relatively low d.c. current of 0.2 mA is used to detect the magnetization state. To eliminate the joule heating, we set a 5 s-time delay between the switching and detection operations. As shown in Fig. 3C, in the presence of a positive in-plane magnetic field of $H_x$ = 350 Oe, the magnetization is switched to the +*z* direction (corresponds to a positive $R_{xy}$) when the magnitude of current pulses reaches above 7.2 mA, while the +*z* to -*z* magnetization switching occurs at a negative current of -6.5 mA. When $H_x$ is switched to -350 Oe, the anti-clockwise magnetization switching with respect to the electric current is changed to clockwise. The similar switching behaviors occur at other $H_x$ values and the critical switching current density slightly increases with the decrease of in-plane magnetic fields (Fig. S4). The chirality of magnetization switching with respect to the in-plane magnetic field and electric current is consistent with the spin Hall angle in Pt, implying the contribution of SOTs from Pt.

Notably, the current-induced change of $R_{xy}$ is almost equal to that induced by the out-of-plane magnetic field, which reveals the performance of full magnetization switching by an electric current. This performance is attributed to the combinational effect of low pinning in the magnet, good interface between the magnet and heavy metal and high ST efficiency. To highlight the merits of our FGT/Pt heterostructure, we compared the operation temperature, the magnetization switching ratio and critical switching current density of our SOT device with other reported results in vdW magnets (Fig. 3D). The performance of room-temperature full magnetization switching by an electric current with low critical current density promises the potential application of FGT/Pt heterostructures for low-consumption and reliable spintronic devices. We further investigate the detailed process of current-induced magnetization switching. As shown in Fig. 3C, a noticeable "tail" is observed at high current pulses in the current-indcued switching hysteresis loop: $R_{xy}$ approaches to the demagnetized state with the increase of currents, which is ascribed to the combinational results of the Joule heating effect and SOTs. It also implies that the magnetization switching is thermally assisted, leading to a low critical switching current density (4.8 MA/cm$^2$).

In the conventional SOT-induced magnetization switching picture, an in-plane magnetic field is indispensable to obtain the deterministic switching of perpendicular magnetization (*36–38*). Without this in-plane magnetic field, we cannot observe the magnetization switching performance in the experiment (Fig.S5). However, in real applications, this is detrimental for energy and miniaturization considerations. We further extend the capability of FGT/Pt heterostructures to realize deterministic perpendicular magnetization switching in the absence of a magnetic field. It has been reported that the in-plane magnetic field can be replaced by the asymmetric device geometry that breaks the mirror symmetry (*39–42*). Taking advantage of the layered structure and flexibility in vdW magnets, we designed an asymmetric device geometry in which the heavy metal Pt covers only one side edge of the magnet FGT (Fig. 4A). In the experiment, the FGT flake sample was placed at a 20° tilt angle with respect to the sputter holder plane. The sample stage didn't rotate during the deposition of the Pt layer, while starting rotating during the deposition of the Ru capping layer. Thus one side of the FGT nanoflake is facing the Pt source and closer toward the incoming sputter flux. As a result, one side edge of the FGT flake is covered by Pt and a wedge of Pt layer is formed on the FGT flake, as illustrated in Fig.4A. The SEM images confirms the coverage situation of Pt on the edges of the FGT flake toward and back to the incoming flux direction (Fig. S6). This geometry breaks the mirror symmetry of the hall device with respect to the *xz* plane, supporting the realization of field-free switching of perpendicular magnetization.

By applying electric currents through the asymmetric FGT/Pt heterostructure, spin currents with both *x*- and *z*-axis polarization are injected from Pt, leading to the magnetization switching. As shown in Fig. 4B, the magnetization of the device can be switched by an electric current even without the assistance of a magnetic field. To eliminate the influence of the remanent magnetic field, all the measurements were conducted with the sample holder placed outside of the electromagnet and the magnetic field (measured by Hall probe) is <0.5 Oe. The positive (negative) current pulse favors the magnetization ⊙ (⊗), which is consistent with the spin polarization in Pt. The field-free switching is further confirmed by the half loop with current sweeping from zero (Fig. 4C). The device is fully saturated under a preset magnetic field of ±1.5 kOe. The current-induced magnetization switching is around 45% which is agreed with the switching ratio of as-grown devices shown in Fig. 4B. The partial magnetization switching can be attributed to the formation of multi-domain state in the FGT flake. As illustrated in Fig. 4D, at the left side of the FGT/Pt heterostructure that is close to the Pt-covered edge, spin currents of the polarization pointing to the *x*- and *z*-axis are injected into adjacent FGT since Pt covers both the top and the left surface of the FGT flake. The combination of *x*- and *z*-axis polarized spin currents can switch the perpendicular magnetization with the resulting toques $\tau \sim \boldsymbol{m} \times (\boldsymbol{m} \times \boldsymbol{x})$ and $\boldsymbol{m} \times (\boldsymbol{m} \times \boldsymbol{z})$. In contrast, at the right side without Pt covering the edge, it only has *x*-axis polarized spin currents and the perpendicular magnetization cannot be deterministically switched in the absence of magnetic field. As a consequence, we observe a field-free switching of partial magnetization in the asymmetric device owing to the contribution from these two regions. By further reducing the size of the FGT magnet to the single-domain regime, it is possible to achieve deterministic and full field-free magnetization switching. Note that this methodology to obtain field-free switching performance is unique for vdW magnets. The metallic magnetic film usually needs to be very thin in order to sustain good perpendicular magnetic anisotropy and hence the modification of Pt coverage on its edge is technically difficult.

**Conclusion**

In summary, we have observed a high current-induced ST efficiency in FGT/Pt heterostructures which is able to modulate the coercive fields by exploiting the non-reciprocal domain-wall propagation and induces full magnetization switching at room temperature. Furthermore, taking advantage of the geometrical flexibility of the vdW materials, we design an asymmetric device by covering one side edge of the FGT flake with Pt and achieve field-free magnetization switching with the additional out-of-plane polarized spin currents. Our approach offers the opportunity to construct efficient, reliable and scalable vdW spintronic devices. Whereas our proof-of-concept demonstration highlights the potential of vdW materials and their compatibilities with state-of-art spintronic techniques, more exciting work can be investigated to boost the competitiveness of vdW magnets via simultaneously achieving full and field-free switching based on all vdW room-temperature spintronic devices.

**Materials and methods**

**Growth of FGT bulk crystal**

High-quality FGT single crystals were grown by self-flux method. Iron powders (Aladdin, 99.95%), gallium lumps (Aladdin, 99.9999%), and tellurium powders (Macklin, 99.99%) were weighed and mixed in a stoichiometric ratio 1:1:2. The mixture was sealed in an evacuated quartz tube, heated to 1273 K with 6 K/min, and held for 24 h for solid reactions. Then the temperature was quickly decreased down to 1153 K within 1 h followed by a slow cooling down to 1058 K within 95 h. During the cooling period of the growth, the residual Te fluxes were removed by a centrifugation process and single crystals were yielded with a typical size of 2 × 2 × 0.2 mm and a cleavable layer in the *ab* plane.

**Characterization of FGT crystal and device**

The characterization of the morphology and crystal structures of FGT crystals were performed by optical microscopy (Olympus), powder X-ray diffraction (X'Pert Pro MPD diffractometer with Cu Kα radiation: λ = 1.5418 Å) and Raman spectroscopy (Horiba, XploRA PLUS with excitation light of 532 nm). The thickness and surface topography were measured by atomic force microscopy (AFM, MFP-3D Origin, Oxford Instruments). The cross-section microstructure was investigated using a field-emission transmission electron spectroscopy (TEM, JEM-F200). The TEM specimen was prepared with a focused ion beam system (FEI, Helios 5 CX). Pt was deposited in sequence for protecting samples from Ga ion implantation.

**Device fabrication and measurement**

To prepare the FGT/Pt device, FGT flakes (thickness ranges between 10 to 30 nm) were first transferred onto $SiO_2$/Si substrates using mechanical exfoliation in the glovebox enclosure with inert gas atmosphere. Then the Pt (7 nm) layer was deposited onto the FGT flake using DC magnetron sputtering (AJA) with a low power of 20 W to eliminate the damage of the FGT/Pt interface by the bombardment of heavy Pt atoms. The base pressure for the sputtering was <5×10$^{-8}$ Torr. The Pt layer was further patterned into a Hall bar geometry using UV lithography and Ar ion milling, with the FGT flake located under the cross centre of the Hall bar. The width of the Hall bar is 5 μm. Different from the symmetric device as described above, the asymmetric device used for field-free switching measurement was fabricated with the FGT flake sample placed at a 20° tilt angle with respect to the sputter holder plane. The sample stage didn't rotate during the deposition of the Pt layer (~5 nm), while starting rotating during the deposition of the Ru capping layer (2 nm).

All the room-temperature magnetotransport properties were performed on a home-made electromagnet. Keithley 2400 source meter and 2182 nanovolt meter were used to apply DC currents and to collect the hall resistances, respectively. The magnetic properties of the bulk and the low-temperature magnetotransport properties of the devices were performed in the vibrating sample magnetometry

(VSM) of a Physical Property Measurement System (PPMS, DynaCool, Quantum Design) system which can provide the environment of high magnetic fields and varied temperatures.

**Figures and figure captions**

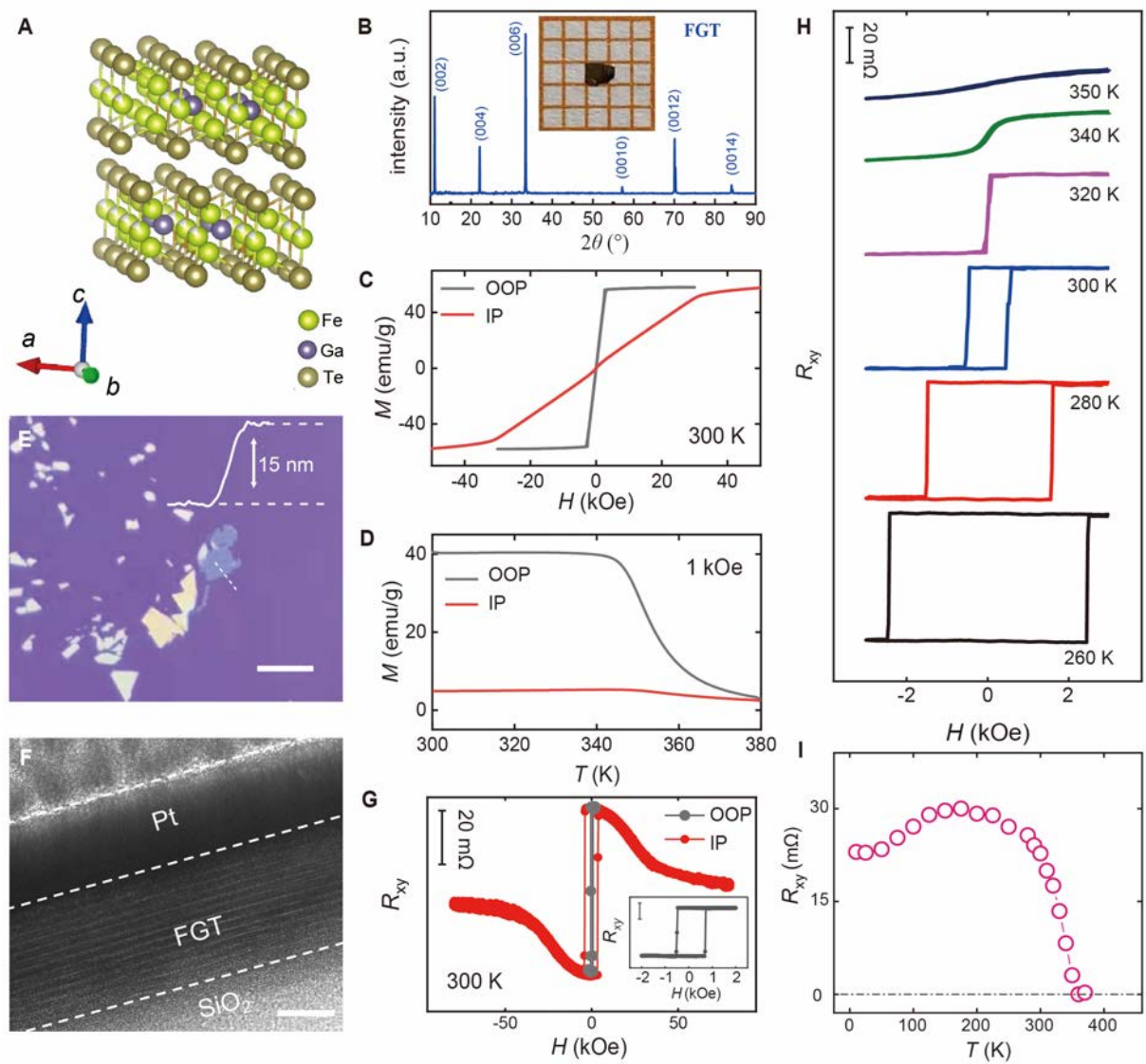

**Figure 1. Structural and magnetic properties of the FGT bulk crystal and FGT/Pt heterostructure.** (**A**) Schematic view of the crystal structure of FGT. (**B**) XRD diffraction pattern and the optical image (inset) of the as-grown FGT bulk crystal. The grid size in the image is 1 mm. (**C**) Magnetization vs. magnetic field (*M-H*) curves of the FGT bulk crystal with out-of-plane (OOP) and in-plane (IP) magnetic fields. (**D**) Magnetization vs. temperature (*M-T*) curves of the FGT bulk crystal with a fixed OOP and IP magnetic field of 1 kOe. (**E**) Optical image of the exfoliated FGT flakes. The line profile on a FGT flake (marked by the white dashed line) shows the thickness of 15 nm. Scale bar: 10 μm. (**F**) Cross-sectional HR-TEM image of the FGT/Pt heterostructure on a Si/SiO$_2$ substrate. Scale bar: 5 nm. (**G**) AHE hysteresis loops of the FGT/Pt heterostructure with OOP and IP magnetic fields. Loops with the small OOP magnetic field is shown in inset. (**H**) AHE hysteresis loops measured at different temperatures with OOP magnetic fields. (**I**) Measured Hall resistance vs. temperature curve of the FGT/Pt heterostructure.

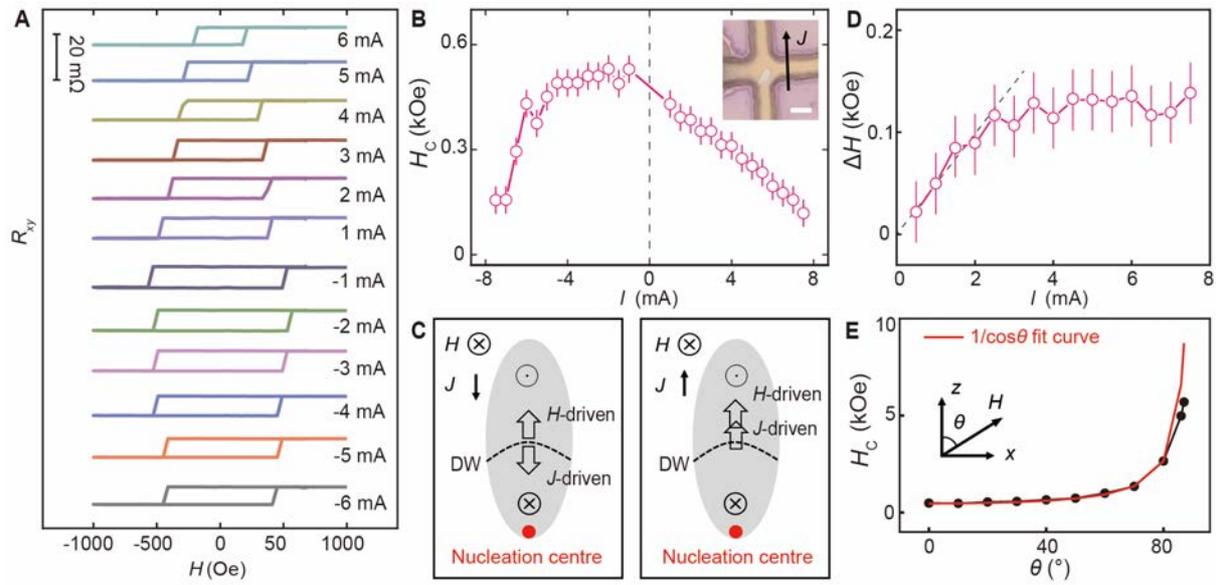

**Figure 2. Spin torque (ST)-induced non-reciprocal modulation of coercive fields in the FGT/Pt heterostructures.** (**A**) Hysteresis loops of $R_{xy}$ measured with different applied currents. (**B**) Coercive field $H_C$ vs. applied current curve in the FGT/Pt heterostructure. The optical image of the device is shown in inset and the direction of applied current is indicated. Scale bar: 5 μm. (**C**) Schematics illustrating the process of domain nuclueation and domain-wall propagation in an elongated FGT flake driven by magnetic fields and current-induced STs. The directions of magnetic fields and applied currents are indicated as well as the their driving forces on the domain walls. (**D**) Effective magnetic field vs. applied current curve in the FGT/Pt heterostructure. (**E**) Angular dependence of $H_C$ obtained from the hysteresis loops.

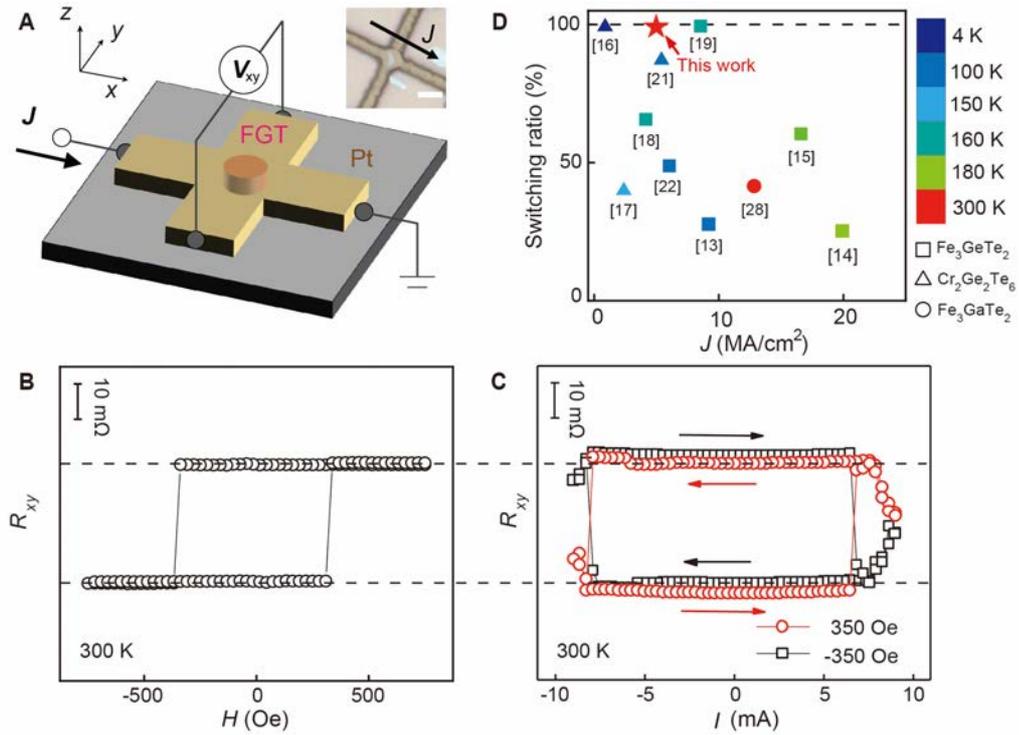

**Figure 3. Current-induced full magnetization switching in the FGT/Pt heterostructure.** (**A**) Schematic illustrating the transport measurement layout. Inset: the optical microscopic image of the measured device. The direction of applied current is indicated. Scale bar: 10 μm. (**B**) Hysteresis loop of $R_{xy}$ driven by magnetic fields along the $z$ axis. (**C**) Hysteresis loops of $R_{xy}$ driven by electric currents with an in-plane magnetic field of ±350 Oe. The arrows indicate the current sweeping direction. (**D**) Comparison of the critical current density, switching ratio and operation temperature (represented by the color of the symbols) of the magnetization switching in vdW magnets.

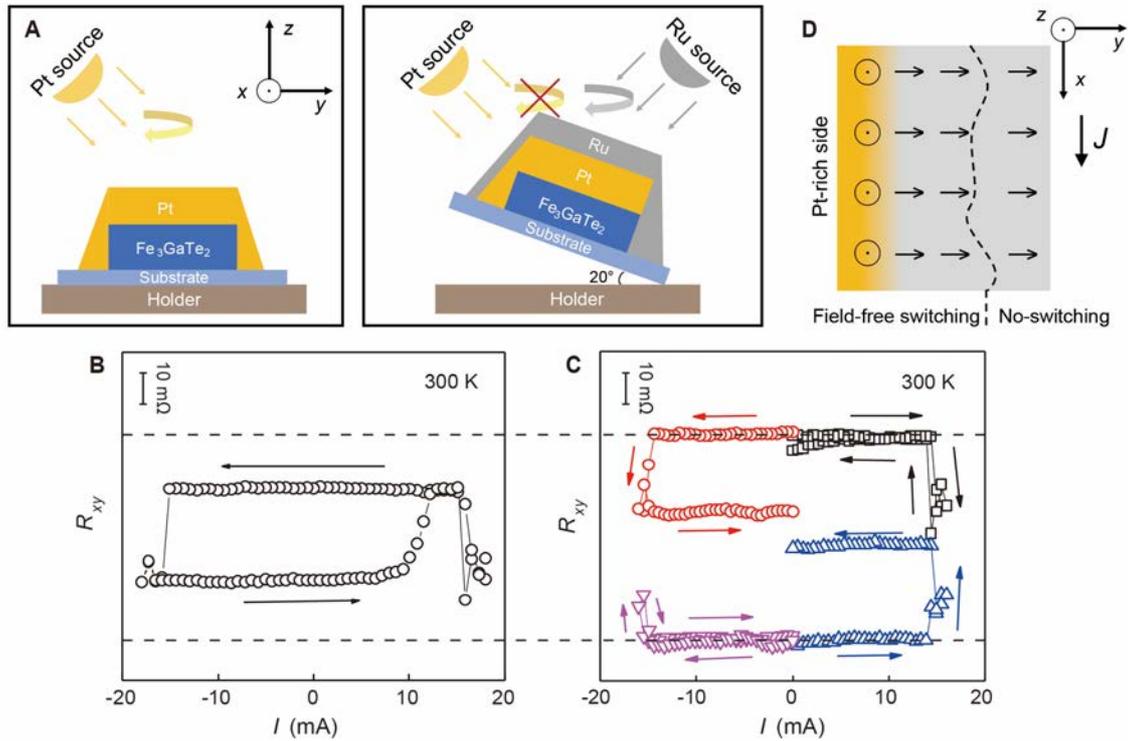

**Figure 4. Field-free magnetization switching of FGT/Pt heterostructures by geometrical design.** (**A**) Schematic illustrating the fabrication process of the symmetric (left) and asymmetric (right) the FGT/Pt device. (**B**) Hysteresis loop of $R_{xy}$ driven by electric currents in the absence of magnetic fields. The dashed lines correspond to the $R_{xy}$ at saturated magnetization states. The arrows indicate the current sweeping direction. (**C**) Four-part half hysteresis loops of $R_{xy}$ driven by electric currents sweeping from zero to positive or negative currents after being fully saturated under a preset OOP magnetic field. The arrows indicate the current sweeping direction. (**D**) Schematics illustrating the field-free magnetization switching in the asymmetric FGT/Pt device. The polarizations of injected spin currents are indicated.